\documentclass[10pt,conference]{IEEEtran}
\IEEEoverridecommandlockouts

\usepackage{amsmath,amssymb,amsfonts}
\usepackage{algorithmic}
\usepackage{graphicx}

\usepackage[font=footnotesize, labelfont=bf]{caption}
\usepackage[font=footnotesize, labelfont=bf]{subcaption}
\usepackage{subcaption}

\usepackage{textcomp}
\usepackage{xcolor}
\usepackage{booktabs}  
\def\BibTeX{{\rm B\kern-.05em{\sc i\kern-.025em b}\kern-.08em
    T\kern-.1667em\lower.7ex\hbox{E}\kern-.125emX}}

\usepackage{multirow}

\begin{document}

\title{Graph Integrated Transformers for Community Detection in Social Networks\\

}

\author{\IEEEauthorblockN{Name}
\IEEEauthorblockA{\textit{\ Department} \\
\textit{Organization}\\
City, Country. \\
Email: }
\and
\IEEEauthorblockN{Name}
\IEEEauthorblockA{\textit{\ Department} \\
\textit{Organization}\\
City, Country. \\
Email: }
}

\author{\IEEEauthorblockN{Heba Zahran}
\IEEEauthorblockA{\textit{\ School of Information Technology} \\
\textit{Carleton University}\\
Ottawa, ON, Canada \\
hebazahran@cmail.carleton.ca}
\and
\IEEEauthorblockN{M. Omair Shafiq}
\IEEEauthorblockA{\textit{\ School of Information Technology} \\
\textit{Carleton University}\\
Ottawa, ON, Canada \\
omair.shafiq@carleton.ca}
}

\maketitle

\begin{abstract}
Community detection is crucial for applications like targeted marketing and recommendation systems. Traditional methods rely on network structure, and embedding-based models integrate semantic information. However, there is a challenge when a model leverages local and global information from complex structures like social networks. Graph Neural Networks (GNNs) and Transformers have shown superior performance in capturing local and global relationships. In this paper, We propose Graph Integrated Transformer for Community Detection (GIT-CD), a hybrid model combining GNNs and Transformer-based attention mechanisms to enhance community detection in social networks. Specifically, the GNN module captures local graph structures, while the Transformer module models long-range dependencies. A self-optimizing clustering module refines community assignments using K-Means, silhouette loss, and KL divergence minimization. Experimental results on benchmark datasets show that GIT-CD outperforms state-of-the-art models, making it a robust approach for detecting meaningful communities in complex social networks.
\end{abstract}


\section{Introduction}
Social networks represent the connections and interactions between individuals or entities, often modelled as graphs where nodes represent entities and edges represent relationships. Community detection in social network analysis identifies groups of closely connected nodes, distinguishing them from other less-connected groups \cite{community_book_2016}. This process plays a vital role in extracting the underlying patterns and structures in complex networks, providing insights into user behaviour and preferences \cite{Probabilistic_community_2023}\cite{influence_similarity_2024}. Users within the same community often share similar interests, which can be leveraged in different applications, e.g., recommendation systems. By detecting communities, businesses can target specific groups with personalized marketing campaigns to reach out to users more likely to be interested in their service\cite{Probabilistic_community_2023}.

The research on community detection within social networks focuses on assigning each user node to exactly one community or focusing on overlapping communities. Traditional approaches to community detection rely heavily on network structure as they focus on user interaction patterns, while deep learning methods leverage node features, graph structures, and embeddings for better community detection. Generally, techniques that incorporate nodes' features rely on learning and extracting the node embeddings to be utilized in exploring the communities in the network. Node2Vec and Deepwalk \cite{node2vec_2016} methods integrate random walk to generate node sequences as sentences for learning embeddings and are considered flexible methods with careful tuning but computationally expensive with dense networks. Graph Neural Networks (GNNs) aggregate and propagate information across the graph to learn node representations. It can be utilized alone and combined with other techniques to enhance community detection task. \cite{unsupervised_2022} utilizes Graph Convolution Networks (GCN) to improve the efficiency and accuracy of detecting communities within social media data based on nodes' features and connections. \cite{actor_critic_2023} utilizes Deep Reinforcement learning (DRL) and Graph Attention Network (GAT) for detecting communities in dynamic social networks. \cite{node_embedding_approach2023} focuses on reducing the computation time by incorporating node2vec and GCN to learn the feature representations of nodes and classify them into communities based on high similarity values.

\paragraph{Challenges and present work} While GNN-based models excel at aggregating information from local neighbourhoods, they may struggle to capture long-range dependencies in large and complex networks. Social networks exhibit local patterns (e.g., friend relationships) and global patterns (e.g., influencers shaping large-scale interactions), and that requires methods seamlessly integrating local neighbourhood aggregation with global context to identify non-obvious community structures. Additionally, real-world social networks are often Heterogeneous Information Networks (HINs), including different types of nodes and edges, and need models to handle rich features for detecting communities effectively. Therefore, there is a pressing need for models that not only model local and global information but also accommodate the diversity of heterogeneous social networks for accurate community detection.

Transformers networks \cite{vaswani2017attention} achieved great success in many artificial intelligence fields using self-attention mechanisms that compute globally, allowing the model to capture long-range dependencies and interactions. The demonstrated limitations motivate us to explore hybrid frameworks combining GNNs with transformer architectures for community detection. Besides the GNNs that excel in capturing the local structure of a graph, we need to integrate a method for a global structure that can handle HIN. In this paper we propose Graph Integrated Transformer for Community Detection (GIT-CD) to investigate the feasibility of incorporating the Transformer architectures with a GNN, to learn graph structure and heterogeneity simultaneously with refining cluster assignments during the learning process in an end-to-end manner. In particular, GIT-CD is a semi-supervised learning model that employs the architecture of GNN and the encoder-based Transformer while incorporating heterogeneity in the standard multi-head attention, allowing the model to capture meaningful relationships between different types of nodes in a graph. The learning objective is to jointly optimize clustering and classification loss to enhance the model's ability to generalize from labelled data while uncovering meaningful structures in the graph through unsupervised clustering.

\paragraph{Contributions} 
We investigate the application of Graph Transformer Networks (GTNs) on HINs for learning community detection. 

In summary, our contributions are: (1) We integrate the Transformer architecture with GNNs to enhance the learning of global structures in complex networks while improving scalability by leveraging the parallel processing capabilities of Transformers for better handling of large-scale graphs. 
(2) We introduce a dynamic multi-head attention mechanism that allows node-type-specific attention to optimize the Transformer-based global attention mechanism to ensure computational efficiency on large-scale networks. (3) We incorporate a self-optimizing clustering module that refines community assignments during training by integrating K-Means clustering, KL divergence minimization, and silhouette loss. 

The remainder of this paper is outlined as follows: Section 2 reviews the related works. In Section 3 presents the proposed methodology and model. Section 4 discusses experimental settings and results. Finally, Section 5 concludes the study and suggests future directions.
\section{Related Work}
Community detection has been extensively studied using various methods. This section provides an overview of GNN-based models and an overview of transformer networks..

\subsection{GNN-based Models for Community Detection}
GNNs are deep learning models designed to operate on graph-structured data. They are suitable for community detection as they have successfully learned node representations through message passing and aggregation. GNNs enabled deep graph clustering (DGC) methods to learn cluster assignments from semantic and topological information based on prior knowledge of the number of communities. 



Recent innovations add modules to GNNs to preserve the dependency and similarity information for community detection tasks. The JGE-CD \cite{JGE-CD_2019} is an unsupervised model that integrates a community detection module based on modularity with a Graph Convolutional Networks (GCN) encoder\cite{GCN2016} and network topology decoder. \cite{CRF_GCN_2019} proposes a Conditional Random Field (CRF) layer for GCN to preserve node similarity. Graph Markov Neural Networks (GMNN) \cite{Gmnn_2019} incorporates a CRF to model the joint distribution of labels across nodes. MRFasGCN\cite{MRF_GCN_2019} is an end-to-end deep semi-supervised learning method that integrates GCN and Markov Random Fields (MRF) for modelling label dependencies. GUCD \cite{GUCD_2021} is a GCN-based approach for unsupervised community detection that incorporates the network modelling method of MRFasGCN in the autoencoder framework and employs a dual decoder to reconstruct network structures and node attributes separately. AC2CD \cite{actor_critic_2023} is an Actor-Critic framework that utilizes Graph Attention Networks (GAT) \cite{GAT_2017} and Deep Reinforcement Learning (DRL) for detecting communities in dynamic social networks. \cite{Masked_Dual_Graph_Autoencoder_2024} uses a GAT and a masked graph to obtain a final hidden representation of the graph and learns graph information representations through a masking operation, respectively. The two graph representations are fused into a final embedding for graph clustering with a self-optimizing clustering module. For overlapping communities, \cite{Overlapping_community2025} is a framework integrating a GAT. \cite{sun2021} integrates a GNN encoding method with a multi-objective optimization to ensure that nodes within the same community are densely connected, while minimizing similarity between different communities.

\subsection{Graph Transformer Networks}
The Transformer network is a simple model designed initially for sequence modelling. It relies on self-attention mechanisms to extract the global dependencies between the inputs and the outputs. The original Transformer consists of stacked self-attention (to weight the focus on token embedding) and point-wise fully connected layers for both the encoder and decoder \cite{Vaswani2017AttentionIAUN}.



Transformers integrated with graphs can take various forms to help address the limitations of GNNs and adopt the capabilities of transformers to process graph structure data. For example, reducing the computation cost when processing large graphs or heterogeneous graphs with complex relations \cite{GraphTrans_2022} \cite{HINormer2023} \cite{HierarchicalGT_2022}. The transformers attention is augmented with the positional encoding to preserve the graph structure while extracting the global relations between the entities \cite{TokenGT_2022} \cite{GraphiT_2021}. In some GTN models, the edge features are injected into the calculated attention scores \cite{GO_Dwivedi2020AGO} \cite{SAN_2021} and utilize Laplacian eigenvectors for position encoding.

While Graph Transformers have shown promise in various tasks, their application to community detection remains underexplored. This research seeks to bridge this gap by integrating GNNs and Transformers into a unified framework for community detection on HIN.

\section{The Proposed Model GIT-CD}
This section provides a detailed description of the GIT-CD model. The model operates on a heterogeneous graph with a full adjacency matrix, introduces dynamic multi-head attention tailored for HINs, and a novel clustering optimization module, combining K-Means, KL divergence minimization, and silhouette loss to refine cluster quality. The GNN module extracts node representations after aggregation for all node types. These representations are then passed to the Transformer block with multi-head attention, which captures high-order semantic relationships within the specified context. The resulting target node embeddings are used for multi-objective learning, incorporating classification and clustering to enhance the model's performance.

\subsection{GNN Module for Graph Embedding}

Utilizes GNNs to capture local structural and semantic information using neural message passing.\hfill\\
\paragraph{Message Passing} Applies the AGGREGATE function that creates a message for each hop based on the neighbours' embeddings and the graph structure. Then, the central node embedding is updated through the UPDATE function, which combines the node embeddings with the neighbour message \cite{GCN2016}. 



\subsection{Transformer Block Module}
Based on the standard encoder-base transformer that employs scaled dot-product attention to capture long-range dependencies and refine node embeddings. The encoder consists of K identical layers, and each layer combines two sub-layers: a multi-head self-attention layer and a position-wise fully connected feed-forward network (FNN). 
A normalization layer follows each sub-layer, with a residual connection around \cite{vaswani2017attention}.

\subsubsection{Dynamic Multi-Head Attention}
Traditional attention mechanisms assume homogeneous input. In our setting, the dynamic of the attention module is based on its ability to be extended to HINs. We incorporate node-type-specific queries, keys, and values, allowing for heterogeneous message passing to learn and process the complex interdependencies between diverse node types. The core idea is to enable each node type to have separate query $Q_t$, key $K_t$, and value $V_t$, ensuring that attention weights are learned differently for different node types. The attention score is computed intra-type (e.g. author-to-author) and inter-type (e.g. author-to-paper). In other words, the module combines self-attention within the same node type is similar to standard self-attention formulation $Score_{v,v}$ \eqref{eq: 5}, and cross-type attention computes the attention between different node $Score_{v,u}$ \eqref{eq: 6}. 
The attention mechanism is repeated $h$ times (number of heads), enabling parallel attention to multiple subspaces of the input simultaneously.

\noindent
\paragraph{Attention scores} The attention score is a combination of the relevance of the target node $v$ to itself and another node type $u$ within a network. For the target node, a self-attention computes the similarity between queries $Q$ and keys $k$ of the same target node type. An across-attention computes the similarity between queries and keys of the target and other node types, respectively. 

\begin{equation}
\label{eq: 5}
    Score_{v,v} = Q_v \cdot K_v^T
\end{equation}

 \begin{equation}
\label{eq: 6}
    Score_{v,u} = Q_v \cdot K_u^T
\end{equation}

These scores are aggregated across specified node types in the network to compute the combined attention scores \eqref{eq: 7}:

 \begin{equation}
\label{eq: 7}
    Attention_Scores_{v,u} = Score_{v,v} + \sum_{u\in T_v}{Score}_{v,u}
\end{equation}
\noindent
\paragraph{Masking} We employ masks to correctly handle dynamic neighbourhood sizes and type-specific interactions. In a HIN, different node types may have different numbers of nodes. The attention computation for all node types is unified under the maximum number of nodes $max_nodes$ for each batch to allow operations like matrix multiplications. The node types with smaller nodes are padded with $zeros$ values. To compute the attention scores correctly, we mask the attention scores created by padded nodes so it does not contribute to the computation as in \eqref{eq: 8}.

 \begin{equation}
\label{eq: 8}
\text{MaskedScore}_{ij} =
\begin{cases} 
\text{Score}_{ij}, & \text{if valid node} \\
-\infty, & \text{otherwise}
\end{cases}
\end{equation}

\noindent
\paragraph{Attention weights}The attention scores are normalized using the softmax function \eqref{eq: 9}. If $Score_{v,u}$ = $-\infty$, then exp$(-\infty) = 0$, ensuring $a_{i,j} = 0$ for invalid nodes/interactions.
\begin{equation}
\label{eq: 9}
    a_{u,v} = \frac{exp(Score_{v,u})}{\sum_k exp(Score_{u,k})}
\end{equation}
\noindent
\paragraph{Aggregation} Using the computed attention weights $\alpha  _{uv}$, the attention-weighted values are computed using the target node $V_v$ value vector \eqref{eq: 10}.

\begin{equation}
\label{eq: 10}
    A_v = \alpha _{u,v} \cdot V_v
\end{equation}
This operation is performed in parallel for all $h$ heads for multi-head attention. The outputs from all heads are concatenated:
\[\ 
A_v = Concat(A_v^{(1)}, A_v^{(2)}, A_v^{(3)}, ..., A_v^{(h)} )
\]
The final representation is trimmed for an iterative learning process.
\subsection{Self-Optimizing Clustering Module}
Combines hard clustering (K-Means) with a soft clustering mechanism using a trainable temperature parameter $t$. The module performs the following:

\noindent
\paragraph{K-Means Clustering} computes hard cluster assignments and cluster centers to initialize the clustering process.
Assign clusters to feature embeddings using the k-means algorithm. The input is a deep feature representation of the target node $A_v$ to be assigned to one of $k-th$ cluster centers $C_k$ minimizes the sum of squared distances between each node's embedding and its closest cluster center \eqref{eq: 11}.

\begin{equation}
\label{eq: 11}
    C = argmin_c\sum _{i=1}^N min_k \parallel A_{v_i} - C_k\parallel^2
\end{equation}

\noindent
\paragraph{Soft Assignment} Computes a soft assignment matrix $Q$ based on cluster centers, where each element $Q_{i,j}$ represents the probability of each node $i$ belonging to each cluster $j$, with a temperature parameter $t$ controlling the softness of assignments. Then $Q_{ij}$ is normalized across clusters \eqref{eq: 12}, \eqref{eq: 13} \cite{soft_2016}. 
\begin{equation}
    \label{eq: 12}
        Q_{ij} = \frac{1}{1 + \biggl(\frac{\parallel A_{v_i} - C_j\parallel^2}{t}\biggl)}
\end{equation}
\begin{equation}
\label{eq: 13}
      Q_{ij} = \frac{Q_{ij}}{\sum_{j=1}^k Q_{ij}}
\end{equation}
\paragraph{Target Distribution} $P$ is computed to refine clustering by amplifying the confident assignments (high $Q_{ij}$) \eqref{eq: 14} \cite{soft_2016}.
\begin{equation}
\label{eq: 14}
    P_{ij} = \frac{\frac{Q_{ij}^2}{f_j}}{\sum_{j=1}^{K} \frac{Q_{ij}^2}{f_j}}
\end{equation}
\noindent
$f_j$ is the sum of all soft assignments for cluster $j$. $P$ is normalized so each sample’s membership probabilities equal $1$.
\subsection{Learning Objectives}
A multi-objective learning where a model is optimized simultaneously with clustering and classification.

\noindent
\paragraph{Clustering Loss} Grouping data into clusters without labels. The Kullback-Leibler (KL) Divergence Loss is used to measure how well the learned feature representations are grouped into meaningful clusters. The process minimizes KL divergence between the target distribution $P$ and the soft assignment matrix $Q$ \eqref{eq: 15} \cite{soft_2016}.
\begin{equation}
\label{eq: 15}
    L_{\text{clustering}} = \frac{1}{N} \sum_{i=1}^{N} \sum_{j=1}^{K} P_{ij} \cdot \log\left(\frac{Q_{ij} + \epsilon}{P_{ij}}\right)
\end{equation}
$\epsilon$: a small constant to prevent numerical instability.

\noindent
\paragraph{Silhouette Loss} is the negative mean silhouette score that evaluates clustering quality by considering intra-cluster cohesion and inter-cluster separation. The primary objective of the training is to encourage nodes within the same community to be embedded closer together while ensuring greater separation from nodes belonging to different communities.
The score $S(i)$ is measured based on \eqref{eq: 17} \cite{Silhouettes_1987} and the loss of $N$ data points as in \eqref{eq: 18}.
\begin{equation}
\label{eq: 17}
s(i) = \frac{b(i) - a(i)}{\max(a(i), b(i))}
\end{equation}
$a(i)$ is the mean distance of $x_i$ to points in the same cluster while $b(i)$ is the minimum mean distance of $x_i$ to points in the other clusters.
\begin{equation}
\label{eq: 18}
L_{\text{silhouette}} = -\frac{1}{N} \sum_{i=1}^{N} s(i)
\end{equation}

\noindent
\paragraph{Classification Loss} The Cross-Entropy Loss is used to measure the error in predicting the correct label for a given input \cite{cross_2023}. 

\noindent
The Total Loss combines classification loss and clustering loss for joint optimization.

\begin{equation}
Total loss = L_{\text{classification}} + L_{\text{clustering}} + L_{\text{silhouette}}
\end{equation}


\section{Experiments}
In this section, we conduct extensive experiments on community detection tasks to evaluate the performance of the proposed GIT-CD.

\subsection{Experimental settings}

\noindent
\paragraph{Datasets} We utilize two public heterogeneous benchmark datasets: a subset of the DBLP computer science bibliography website and a subset of the Internet Movie Database (IMDB) as collected by \cite{blp2020} and downloaded from the Pytorch-Geometric library \cite{pytorch_2019}.



\noindent
\paragraph{Baselines}
To evaluate our model GIT-CD against the state-of-the-art approaches, we consider graph embedding-based models and the standard Transformer network: Metapath2vec \cite{metapath2vec_2017} and MAGNN \cite{Magnn2020} leverage meta-path information. HeCo \cite{HeCo2021}: a contrastive learning framework for HINs. HAN \cite{HaN2019} and GAT \cite{GAT_2017} employ attention mechanisms. HAE \cite{HAE_2021} and URAMN \cite{URAMN2020} integrate local-global optimization techniques. In addition,  GraphConv \cite{graphConv_2019}, SAGEConv \cite{SAGE2017}, and standard Transformer network \cite{vaswani2017attention}.

\noindent
\paragraph{Metrics}
Normalized Mutual Information (NMI) \cite{nmi2002} measures the similarity between the clustering results and ground truth labels, providing a normalized value in the range $[0,1]$. Higher values indicate better alignment.
Adjusted Rand Index (ARI) \cite{ARI1985} counts how many pairs of samples are assigned to the same or different clusters in both the ground truth and predicted labels and adjusts for chance where the score lies in the range [-1,1]. Higher values indicate better alignment.
Silhouette Score \cite{Silhouettes_1987} is a metric to evaluate the quality of clustering results. It measures how each sample is similar to its cluster compared to other clusters. Ranges between [-1,1] and higher scores indicate better-defined clusters.

\noindent
\paragraph{Settings and parameters} We conduct multi-class node classification as an end task on DBLP and IMDB, and then self-optimized clustering is applied. We follow the standard split as in \cite{blp2020}. Accuracy is used as a metric to evaluate the classification performance. NMI, ARI, and Silhouette Score are used to measure the clustering quality. 
For Metapath2vec  we set the walk length to 50, the window size to 7, the walks per node to 5, and 5 negative samples. We test all the meta-paths for all the models
and report the best performance. In terms of other parameters, we follow the settings in their original papers.
GIT-CD has one layer of SAGEGraph for graph embeddings and two blocks of transformers. In training, we utilize a heterogeneous mini-batch graph sampling algorithm—HGSampling \cite{loader_2020}—for efficient and scalable training. We adopted the Adam optimizer and ran experiments on PyTorch. We set the learning rate to $3e-4$, L2 to $5e-4$, the dropout rate to 0.8, and the maximum epoch to 200. epochs. Lazy parameters are initialized by forwarding a single batch to the model. We terminated the training process when the loss failed to decrease for five consecutive. 
The embedding dimension is set as 128 for all experiments, repeated five times, and we report the averaged results. The implementation will be released on GitHub upon acceptance.


\subsection{Results and discussion}
We only focus on detecting disjoint community structures and adopt the NMI, ARI, and ACC to measure the performance of all methods against the ground truth. We report the performance comparison in Table \ref{tb:2} and observe that GIT-CD can outperform all the baselines.

\begin{table}[htbp]
\caption{Quantitative Results on Node Clustering}
\label{tb:2}
\centering
\small  
\begin{tabular}{|c|c|c|c|c|c|c|}
\hline
\textbf{Models} & \multicolumn{3}{c|}{\textbf{IMDB}} & \multicolumn{3}{c|}{\textbf{DBLP}} \\
\cline{2-7}
& \textit{NMI} & \textit{ARI} & \textit{ACC} & \textit{NMI} & \textit{ARI} & \textit{ACC} \\
\hline
Metapath2vec   & 19.74 & 21.16 & 83.21 & 77.68 & 82.83 & 22.47 \\
\hline
URAMN          & --    & --    & --    & 78.46 & 83.99 & 94.15 \\ \hline
MAGNN          & 15.58 & 16.74 & 60.50 & 80.81 & 85.54 & 93.61 \\ \hline
HAN            & 19.87 & 18.66 & 67.83 & 69.65 & 76.41 & 90.51 \\ \hline
HAE            & 32.63 & 31.28 & 58.00 & 77.64 & 79.79 & 93.07 \\ \hline
GraphConv      & 47.05 & 53.11 & 83.21 & 22.62 & 4.06  & 91.37 \\ \hline
GraphGAT       & 28.05 & 28.58 & 74.64 & 69.93 & 76.21 & 90.57 \\ \hline
SAGEConv       & 47.94 & 54.17 & 83.61 & 70.98 & 77.13 & 91.71 \\ \hline
HeCO           & --    & --    & --    & 74.51 & 80.17 & 91.59 \\ \hline
Transformer    & 79.39 & 85.23 & 94.94 & 76.82 & 83.04 & 93.28 \\ \hline
\textbf{GIT-CD} & \textbf{80.15} & \textbf{86.05} & \textbf{95.37} & \textbf{94.31} & \textbf{96.78} & \textbf{98.71} \\
\hline
\end{tabular}
\end{table}

Existing models HIN embedding, such as HAN and MAGNN, often fail to capture higher-order semantic relationships. This limitation impacts the performance of community detection tasks. Our model enables the simultaneous capture of both local and global graph structures.

The standard Transformer can achieve competitive performances against all the GNN-based models with the global attention mechanism between the node embeddings. GIT-CD outperforms all other models, especially the standard Transformer and HAN, by a large margin. This demonstrates the effectiveness of GNN in capturing local structure and heterogeneous-based attention to capture both the underlying contexts and semantics for node representation learning.

On the DBLP dataset, our model achieves an NMI score of $94.31\%$, a $27.54\%$ and $17.49\%$ improvement over HAN and the Transformer, respectively. While the Transformer performed very well on the IMDB dataset, GIT-CD outperformed it by around $1\%$ in ARI and NMI. Metapath2vec perform less than modern approaches like GNNs and Transformers. 

GIT-CD and Transformer model excel in both datasets, but GIT-CD shows superior results in DBLP. This suggests that richer semantic and topological features in DBLP enhanced GIT-CD's ability.

We believe GIT-CD's performance is relevant to cross-type attention. Unlike standard GAT, our model allows different node types to have separate attention transformations and improves heterogeneous node representation learning.

\begin{figure}[htbp]
    \centering

    \begin{subfigure}[b]{0.15\textwidth}
        \centering
        \includegraphics[width=\textwidth]{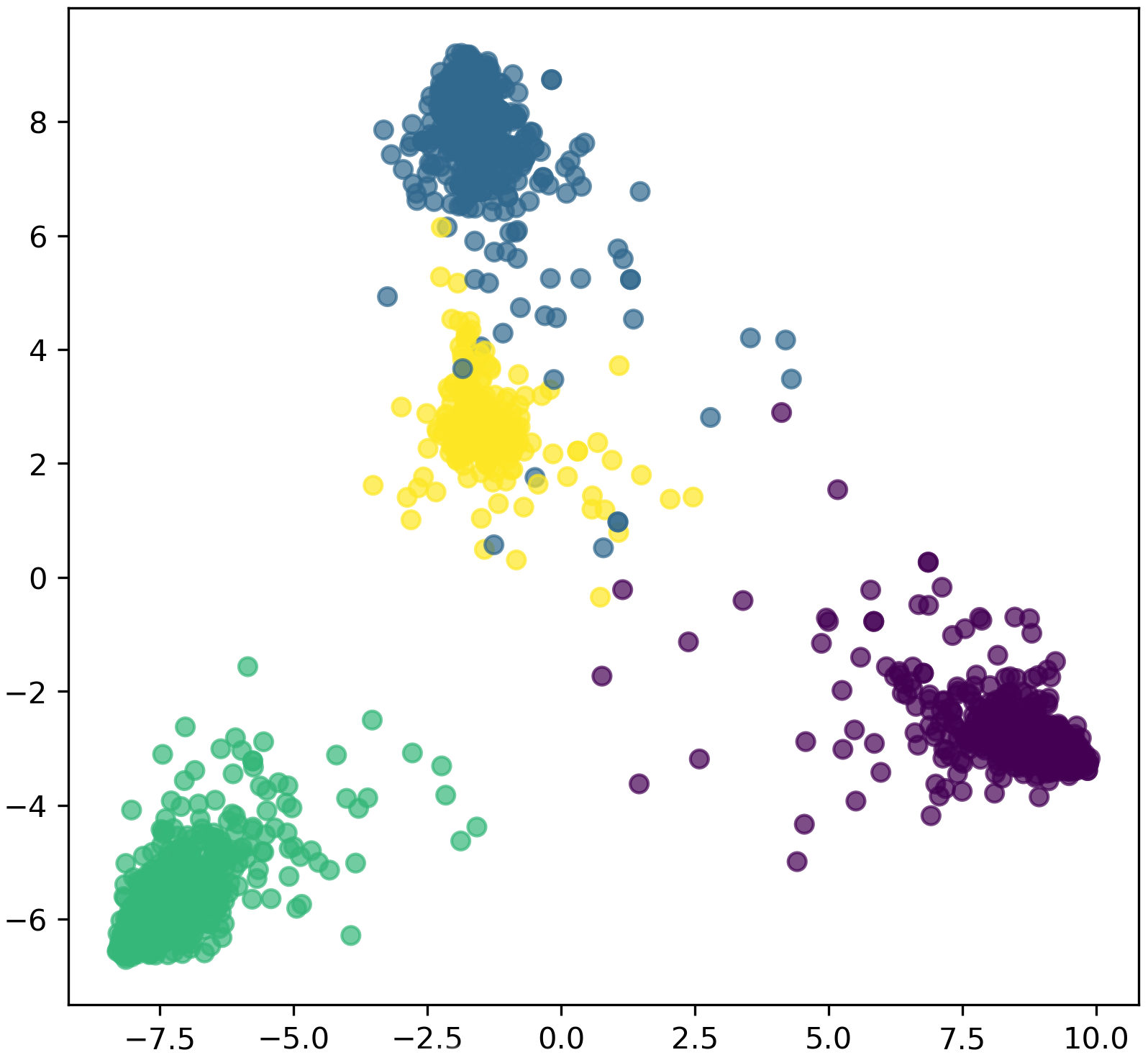}
        \caption{ GID-CD}
        \label {fig:gid}
    \end{subfigure}
    \hfill
    \begin{subfigure}[b]{0.15\textwidth}
        \centering
        \includegraphics[width=\textwidth]{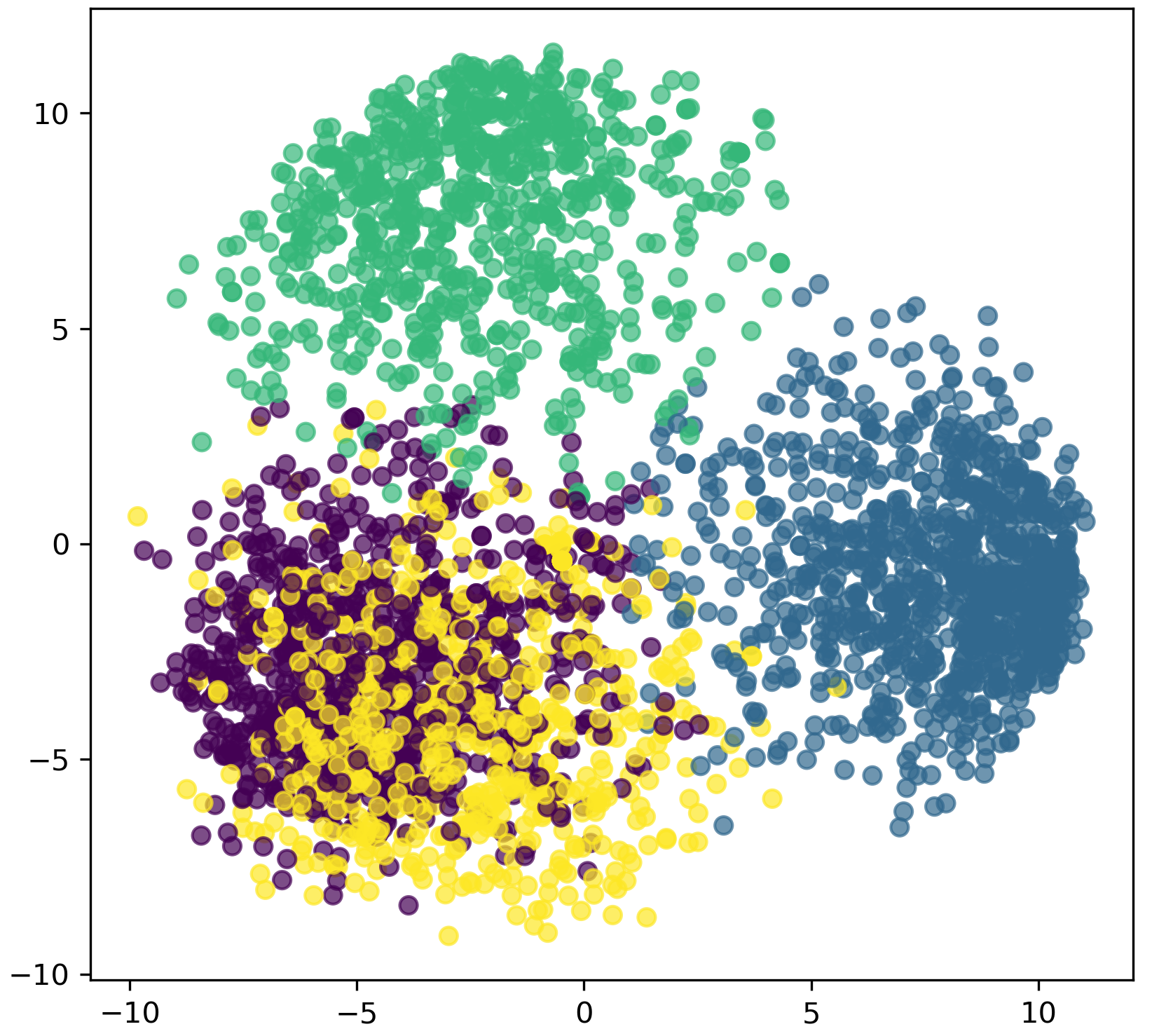}
        \caption{ Transformer}
        \label{fig:transformer}
    \end{subfigure}
    \hfill
    \begin{subfigure}[b]{0.15\textwidth}
        \centering
        \includegraphics[width=\textwidth]{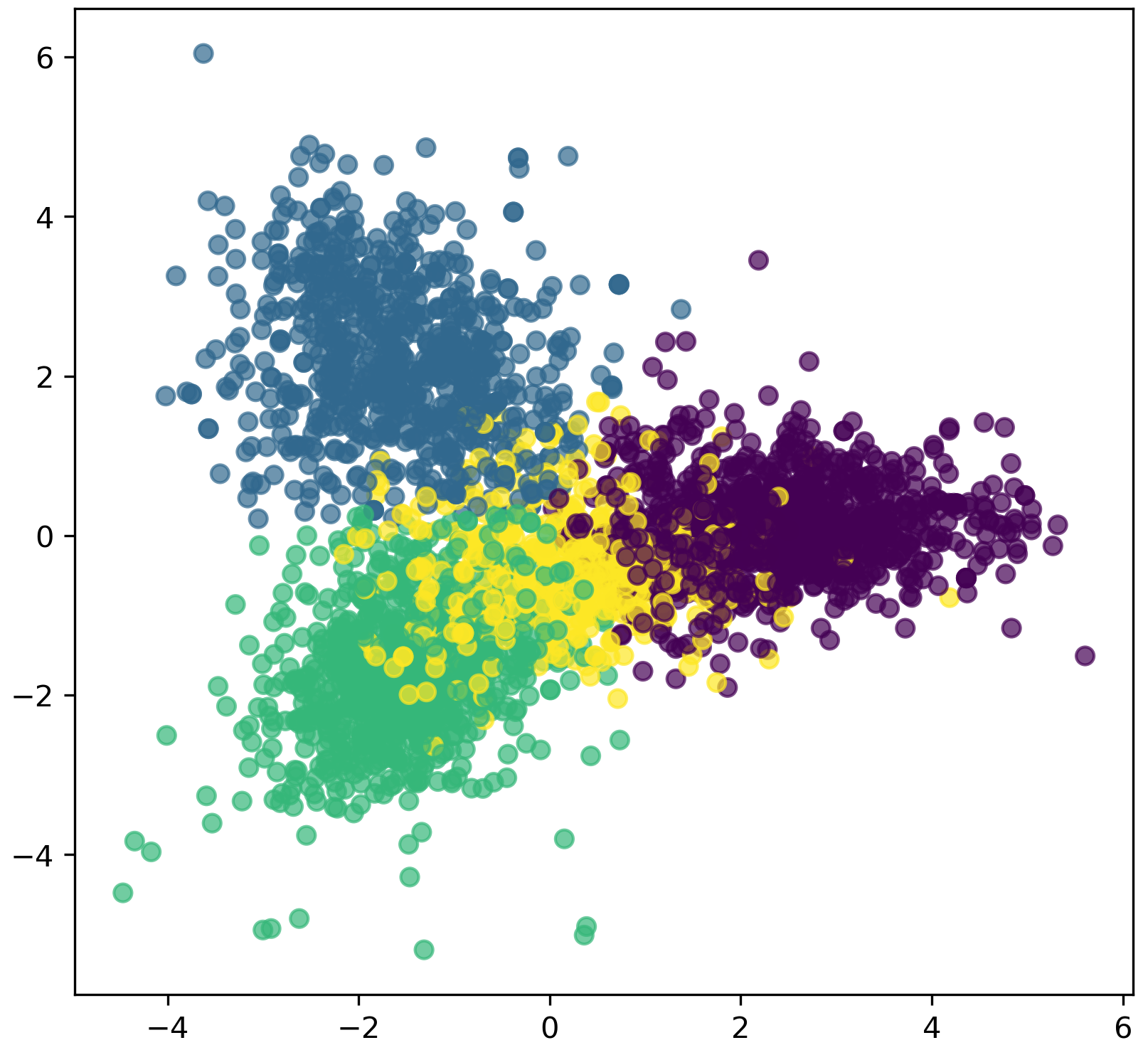}
        \caption{ GAT}
        \label{fig:gat}
    \end{subfigure}
    \hfill
    \begin{subfigure}[b]{0.15\textwidth}
        \centering
        \includegraphics[width=\textwidth]{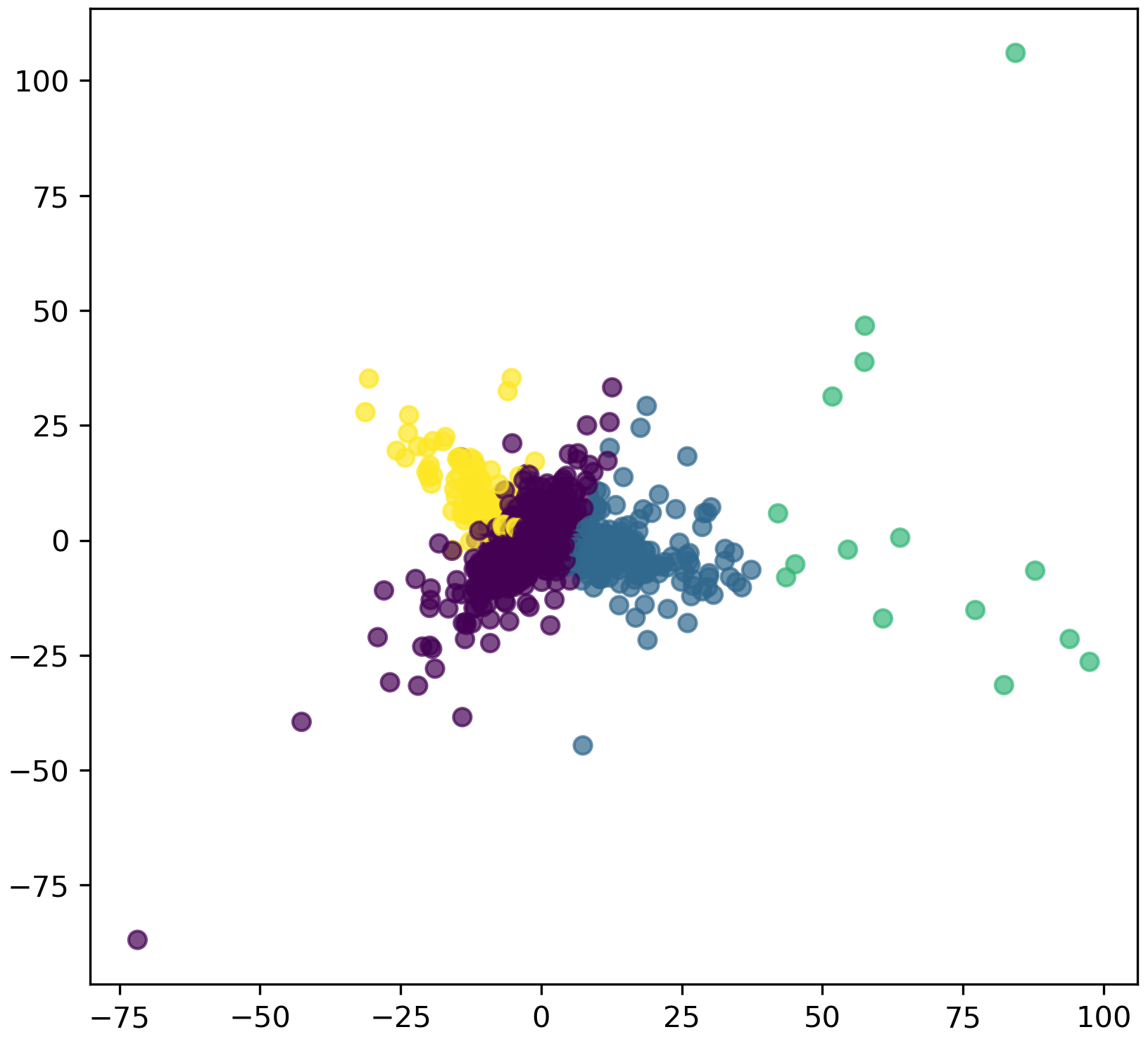}
        \caption{ GCN}
        \label{fig:gcn}
    \end{subfigure}
    \hfill
    \begin{subfigure}[b]{0.15\textwidth}
        \centering
        \includegraphics[width=\textwidth]{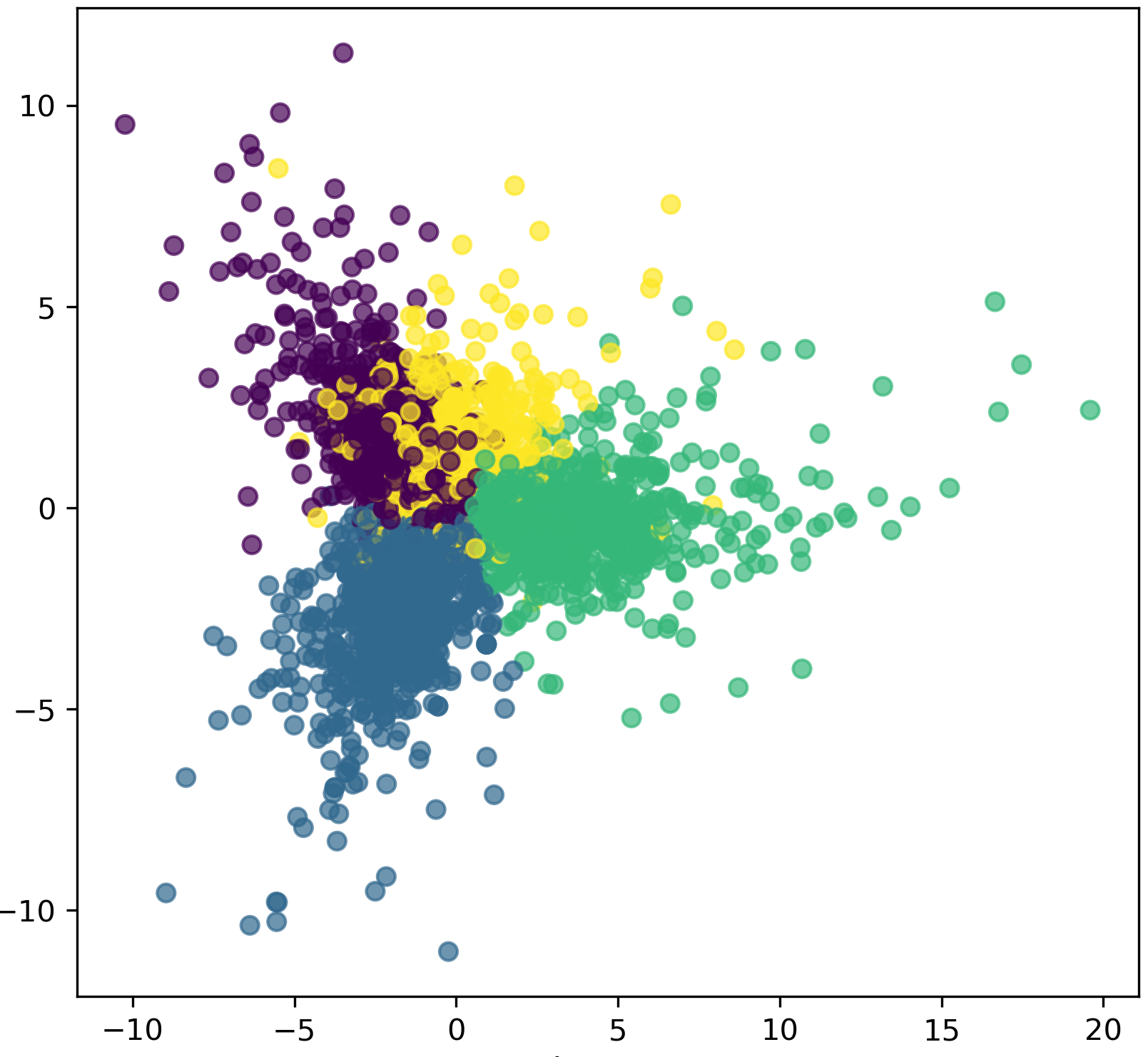}
        \caption{ SAGE}
        \label{fig:sage}
    \end{subfigure}
    \hfill
    \begin{subfigure}[b]{0.15\textwidth}
        \centering
        \includegraphics[width=\textwidth]{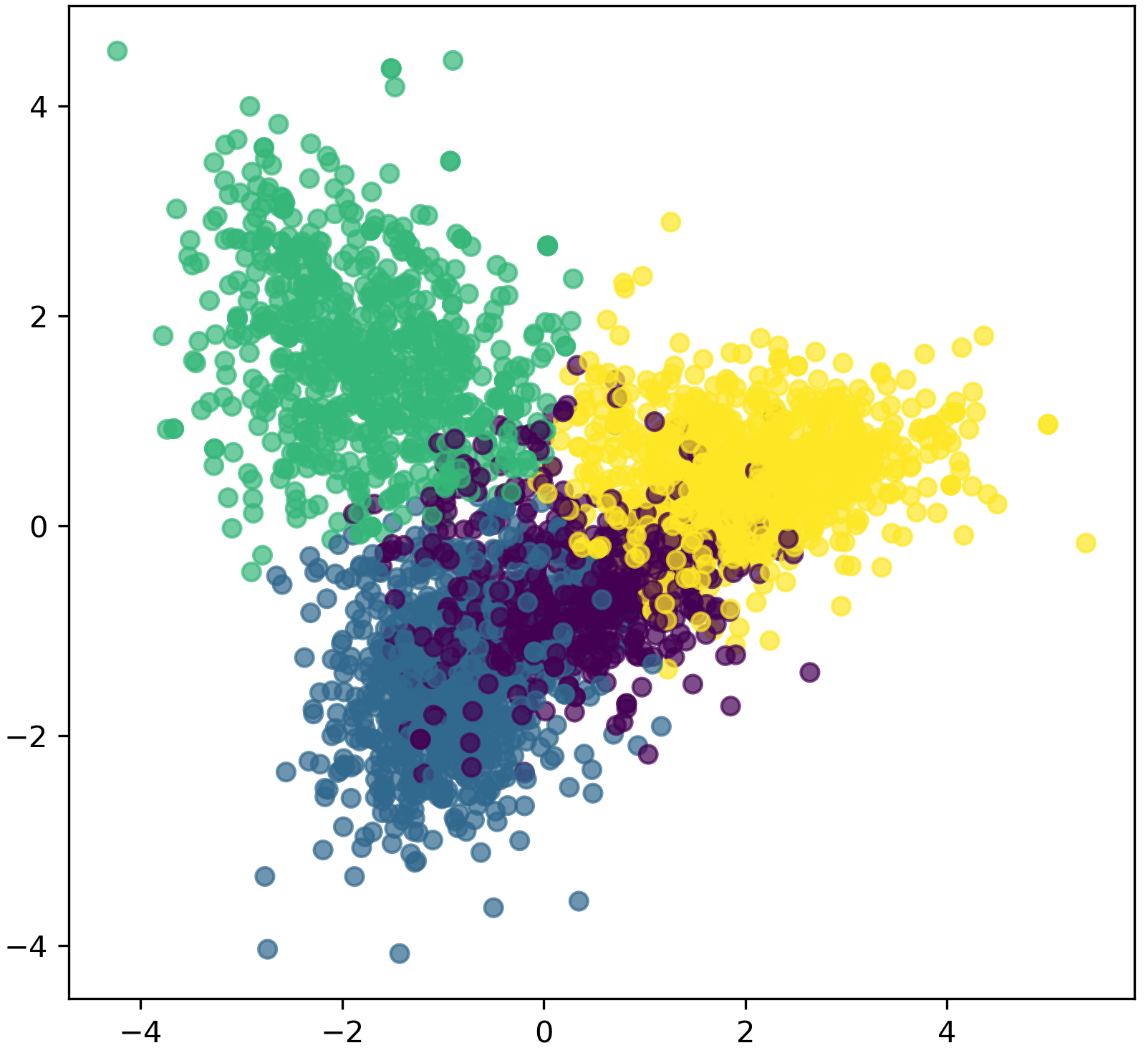}
        \caption{ HAN}
        \label{fig:han}
    \end{subfigure}

    \caption{Visualization embedding on clustered data points of DBLP.}
    \label{fig:embedding}
\end{figure}

As the main purpose of GID-CD is to detect communities, it is essential to discuss the robustness of the model in achieving its main task. The model demonstrates high NMI and ARI scores on both datasets. However, these metrics do not provide insights into the internal quality of the clusters themselves. This is where the Silhouette Score becomes crucial. We report the Silhouette Score comparison between GID-CD and GNN models in Table \ref{tb: 3} and a visual representation as in Fig.~\ref{fig:embedding}. Our model achieved high scores for NMI/ARI and Silhouette Scores. While the other models achieved high accuracy, NMI, and ARI scores, the Silhouette Scores were lower or weak. This suggests that our model produces clusters that align with the ground truth while being internally cohesive and well-separated. The low Silhouette in other models indicates fragmented clusters or clusters that overlap, even though they match ground truth labels.
\begin{table}[htbp]
\caption{Node clustering evaluation based on Silhouette Score}
\begin{center}
\begin{tabular}{|c|c|c|}
\hline
\textbf{Models}     & \textbf{IMDB} & \textbf{DBLP}  \\ \hline
HAN         & { 38.48}                                              & 47.54 \\ \hline
GraphConv   & { 39.54}                                              & 42.39 \\ \hline
GraphGAT    & { 35.53}                      & 46.11 \\ \hline
SAGEConv    & { 39.69}                                              & 38.39 \\ \hline
Transformer & \multicolumn{1}{l|}{{ 72.56}}      & 54.04 \\ \hline
GIT-CD      & \multicolumn{1}{l|}{\ \textbf{92.23}} & \ \textbf{90.87} \\
\hline
\end{tabular}
\label{tb: 3}
\end{center}
\end{table}

\section{Conclusions}
We introduced GIT-CD a hybrid model that integrates GNNs and Transformer-based self-attention mechanisms to enhance community detection in heterogeneous social networks. GIT-CD achieves competitive performance on benchmark datasets by integrating local graph embeddings, global representation, and a self-optimizing clustering module. The self-optimizing clustering module refines community assignments using a combination of K-Means clustering, silhouette loss, and KL divergence minimization, which enables the model to balance supervised classification with unsupervised clustering, improving the quality of detected communities.
We demonstrated that GIT-CD outperforms state-of-the-art models regarding NMI, ARI, and Silhouette Score.


\bibliographystyle{unsrt}

\bibliography{referencias.bib}

\end{document}